# Fully superconducting machine for electric aircraft propulsion: study of AC loss for HTS stator

Fangjing Weng[1*], Min Zhang[1*], Tian Lan[1], Yawei Wang[1] and Weijia Yuan[1]

[1]Department of Electronic and Electrical Engineering, University of Strathclyde, G1 1XQ, Glasgow, UK.

E-mail: min.zhang@strath.ac.uk; wengfangjing@vip.163.com

**Abstract**

Fully superconducting machines provide the high power density required for future electric aircraft propulsion. However, superconducting windings generate AC losses in AC electrical machine environments. These AC losses are difficult to remove at low temperatures and they add an extra burden to the aircraft cooling system. Due to heavy cooling penalty, AC losses in the HTS stator, is one of the key topics in HTS machine design. In order to evaluate the AC loss of superconducting stator windings in a rotational machine environment, we designed and built a novel axial-flux high temperature superconducting (HTS) machine platform. The AC loss measurement is based on calorimetrically boiling-off liquid nitrogen. Both total AC loss and magnetisation loss in HTS stator are measured in a rotational magnetic field condition. This platform is essential to study ways to minimise AC losses in HTS stator, in order to maximum the efficiency of fully HTS machines.

1. Introduction

The power density of current electrical machines is not high enough for advanced propulsion applications in all electric aircraft [1-12]. Comparing to conventional machines with copper windings, superconducting machines can significantly reduce machine volume and weight, thus power density can be significantly increased. The latest High temperature superconductors (HTS) have high critical currents at relatively high operational temperature, which can significantly increase the machine power density. Partially HTS machines based on HTS rotors and copper stators have been developed [13, 14]. However, the cold-rotor-warm-stator design in previously led to large air gaps, therefore reducing the magnetic field. To fully utilize the advantages of HTS windings, fully HTS machines have been proposed to maximize machine power density for applications [10, 15].

In a fully HTS machine, the HTS stator is subjected to a rotational AC magnetic field, as well as an AC transport current. So it generates AC losses, which will potentially increase the size and weight of the machine cooling system. The key challenge of developing fully HTS machine is to minimise HTS winding AC losses. So far, most of the research on HTS tapes and coils focused on transport AC loss or magnetisation loss in a uniform magnetic field [16-22]. There are few studies on the AC loss of HTS in a rotational magnetic field [23]. The US Air force proposed to measure the magnetisation loss in a rotational magnetic field generated by a radial type permanent magnet rotor. But the study doesn't consider the influence of transport currents [10].

Previous studies on the calorimetric method for AC loss measurement of HTS at liquid nitrogen temperature (77 K) have been reported [10], which includes an AC background magnetic field. The system measured the boil-off rate of LN2, the main advantage of using the calorimetric method is that it can measure the total loss from an HTS coil regardless of the phase difference between the applied current and the background field [24, 25]. There is a lack of measurement systems to study AC loss in rotational machines. In this paper, a novel HTS machine platform is proposed based on the calorimetric method in order to measure the total AC loss in a rotational magnetic field. The second chapter demonstrated the whole system design and setup, the third chapter explains system calibration and validation; the fourth chapter illustrated the AC loss measurement results, and the fifth chapter concluded and discussed the results.

## 2. The axial-flux HTS machine platform design

*2.1 Machine design*

To enable the LN2 boil off measurement inside a rotational machine, we have chosen the axial-flux machine design in order to accommodate a measurement chamber for an HTS stator winding. As shown in Figure 1. This machine contains two four-pole permanent magnet rotor discs, two silicon steel back iron and a three-phase HTS stator. Each rotor disc consists of four big NdFeB permanent magnets (100 mm in diameter) sitting on a laminated silicon steel plate. The HTS stator disc is sandwiched between two rotor discs and consists of 6 stator coils, one measurement coil was placed in a separately measurement chamber with liquid nitrogen. The measurement chamber is fully emerged in LN2 to minimise heat transfer. It is connected to a flow meter to measure the flow rate of nitrogen gas. Both rotor and stator are fully emerged in liquid nitrogen during operation.

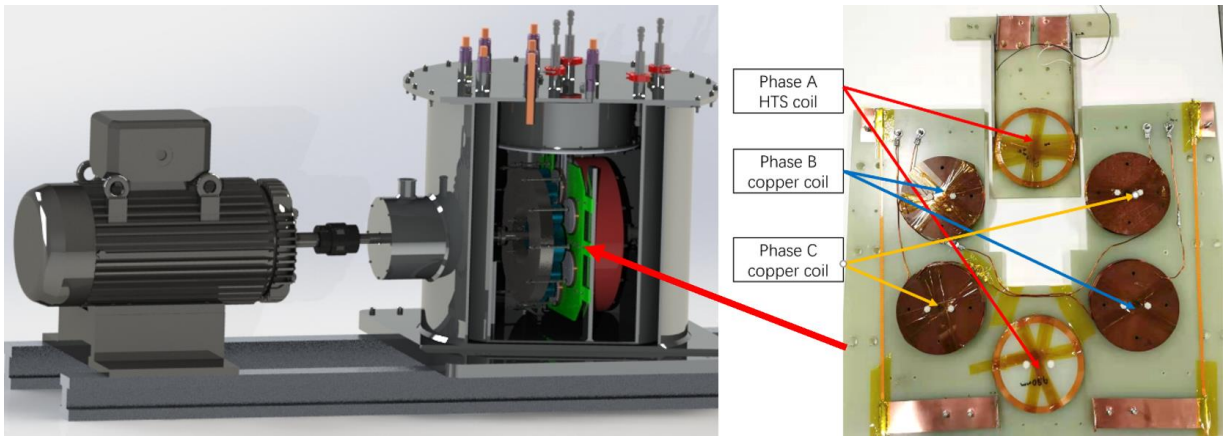

Figure 1. Structure of the measurement chamber

The machine is hosted in a LN2 cryostat with a rotational seal. During the measurement, the rotors are driven by a DC machine to rotate. The HTS stator coil is connected to a load bank, generating three-phase electricity. Only two coils of phase A is made by HTS material as shown in figure 4, phase B and phase C are made by copper coils with the same diameter.

Due to the limit of the bearing, this machine is designed to operate at a speed of 300 RPM, which means a 10Hz three-phase voltage output. A hall sensor was placed across the air-gap in order to measure the magnetic field in the coil center, the peak magnetic flux density of 0.45 T was measured in the center as Figure 2 (a) shown. The superconducting phase A voltage was generated by 2 HTS coils, the voltages of phase B and C consisted of copper coils. Peak voltage of HTS is slightly higher due to the inductance difference between HTS material and copper (copper coil inductance = 526.4µH while HTS coil inductance = 937.4µH). By analysing the spectrum of each phase voltage by FFT calculation, only the 3$^{rd}$ harmonic with a small amplitude exist in stator windings due magnet field distribution as well as non-liner performance of the silicon steel, 3$^{rd}$ harmonic can be removed by star-connection. Thus, this machine can provide a good sinusoidal voltage as Figure 2 (b) shown.

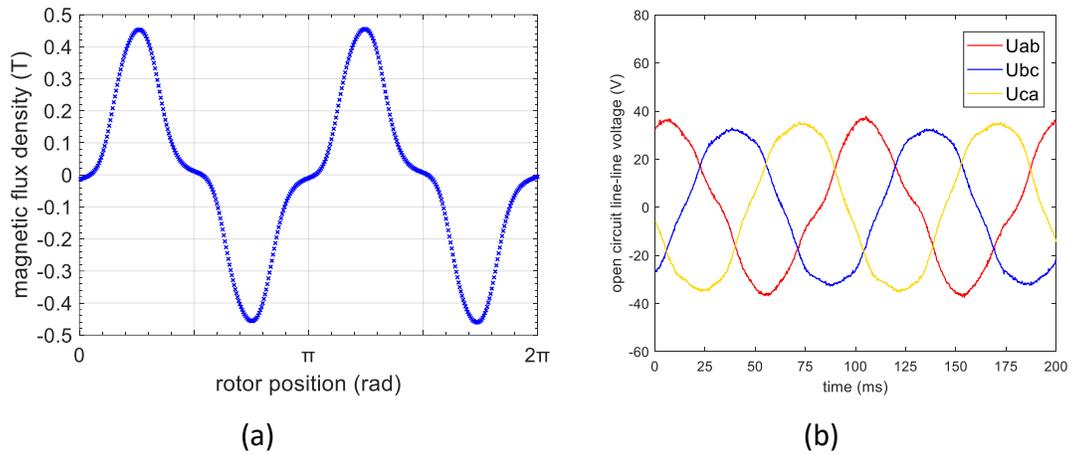

(a)     (b)

Figure 2. (a) Measured magnetic flux field distribution; (b) 3-phase line-line voltages

*2.2 Cryostat design*

To measure the boil-off rate of liquid nitrogen coming from the total AC loss of HTS stator coils, a dedicated measurement chamber was designed, as illustrated in Figure 3. Only one HTS stator coil is placed in the measurement chamber. The measurement chamber and the outer machine chamber are both filled with liquid nitrogen. In an ideal situation, there is no heat transfer between the two chambers. The only conducting component inside the measurement chamber is the HTS stator coil, which makes sure that the boil-off of liquid nitrogen is only due to the total AC loss of the HTS coil.

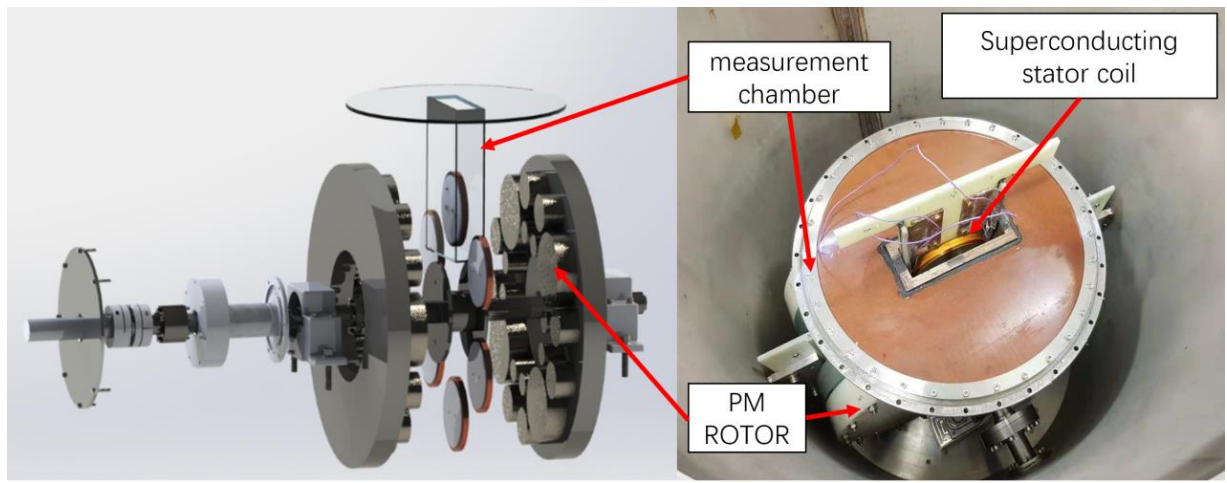

Figure 3. Measurement chamber configuration

The nitrogen gas flow rate is measured by a flow meter (Omega FMA 2710). Theoretically, the latent heat for liquid nitrogen is 160.6 J/mL [26], which equals to 0.25 standard litre per minute for every Watt of power (SLPM/W). Total AC loss of T can be calculated by Equation 1 [19].

$$Q = \int_T \frac{F(t)}{K} dt \tag{1}$$

where $Q$ (Joule) is the total heat produced in the measurement chamber for a duration of T, $F$(t) is the flow rate of nitrogen gas boiled off in the measurement chamber measured by a flow meter. *K=0.256* (Litre/min/Watt) is the flow rate constant. When HTS machine operated in steady state with a fixed frequency *f*, *F(t)* in Equation 1 is simply a constant value with a very small fluctuation, after calibration procedures, we can calculate coil AC loss accordingly. By measuring the nitrogen gas flow rate, calibrate the value of AC loss part, AC loss can be calculated.

*2.3 System setup*

The whole system setup is illustrated in Figure 4, the separately excited DC motor is driven by 2 DC power supplies, the LN2 level in cryostat is observed by 4 PT100 temperature sensors, LN2 boil-off flow rate is recorded by a flow meter, all data were recorded by a NI DAQ system, and HTS machine status including coil voltages , coil currents, motor speed, flow meter data and LN2 level are illustrated by a monitor.

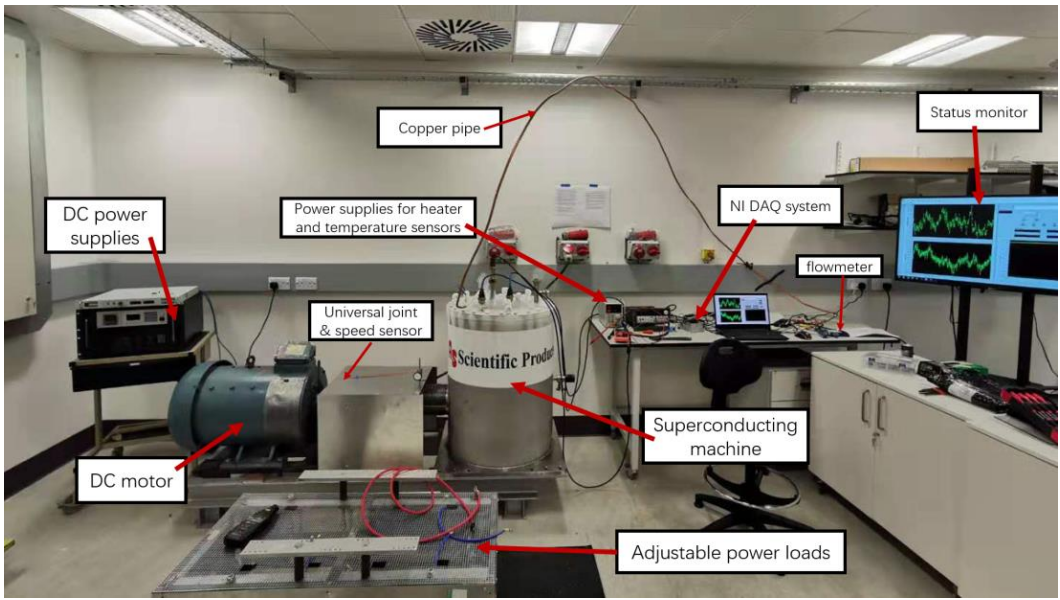

Figure 4. Total HTS machine system configuration

## 3. System calibration and validation

According to the platform setup, total heat input to the measurement chamber consist of the following parts in Equation 2.

$$Q_{total} = Q_{HTS} + Q_{background} + Q_{rotation} + Q_{terminal} \quad (2)$$

$Q_{total}$ refers to the total heat produced in the chamber, $Q_{HTS}$ refers to the heat caused by the HTS coil AC loss, $Q_{background}$ refers to the unavoidable heat leakage in the system, $Q_{rotation}$ refers to the thermal balance condition change when liquid nitrogen was stirring by the rotor and the Joule heat produced by other stator coils, $Q_{terminal}$ refers to the Joule heat cause by copper terminal and solder joint resistance between HTS and copper current leads. These losses in Equation 2 can be quantified by a set of calibration procedures.

In the calibration procedures, the HTS coil in the measurement chamber have been replaced by a very short HTS tape between two copper terminals, and one heater made of Kanthal resistance wire (26.4 ohms) was placed in the system as shown in Figure 5 (a). Figure 5 (b) demonstrate the normal measurement setup with an HTS coil.

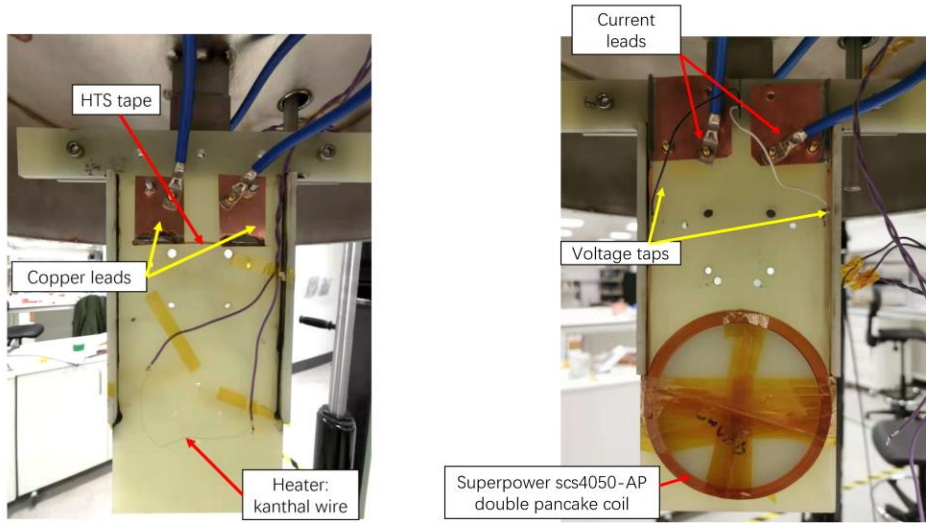

(a) calibration setup, without HTS coil     (b) measurement setup, with HTS coil

Figure 5. Setup in measurement chamber

*3.1 $Q_{background}$ measurement*

Before the calibration starts, liquid nitrogen was fulfilled in both the measurement chamber and the outer cryostat, ensuring the system was cooled down to 77 K. The flow meter (Omega FMA 2710) is connected to the measurement chamber via a long copper pipe. When rotor is static, a background flowrate refers to $Q_{background}$ in Equation 3 was measured, the flow rate was measured for 3000 seconds, the result of flow meter as shown in Figure 6, the background flow of 1.1 SLPM was observed. According to Equation 1 the background heat power is 4.29 W due to heat leakage and environment radiation.

Figure 6 illustrated background flow rate can be regard as a constant value in a short time. $Q_{background}$ can be calibrated out by subtracting the background flow rate.

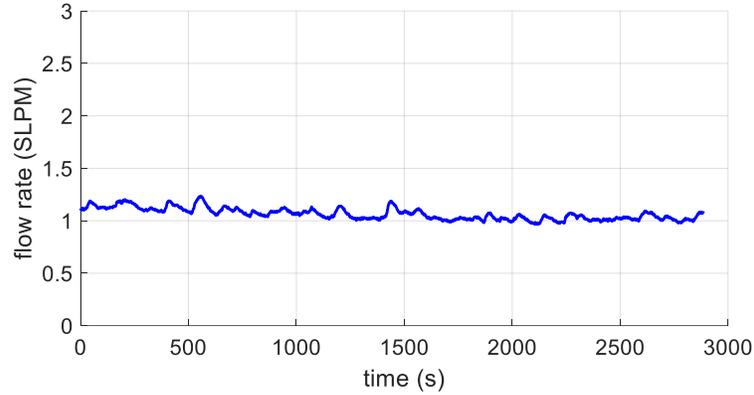

Figure 6. Background flow rate data

*3.2 $Q_{terminal}$ calibration*

As shown in Figure 5 (b), two terminals of the HTS coil were soldered to 2 copper current leads, and then connected to 2 copper bars to fed through in the top flange, so there exists a contact resistance in the solders between the HTS tape and the copper current leads and also a small resistance in the copper bars. When there is a transport current, these is a $Q_{terminal}$. The impact of terminal resistance can be measured by shorting two copper terminals. In this measurement, the HTS coil was replaced by a short HTS tape as shown in Figure 5 (a). A 6 cm HTS tape was soldering between two copper current leads. A DC power supply was connected from the outside and then provides a current from 0 - 80 amps. According to Equation 1, the heat power (watt) can be expressed by Equation 3.

$$I^2 R_{terminal} = \frac{\Delta F}{K} \tag{3}$$

Where *I* is a constant DC current, $R_{terminal}$ is terminal resistance, ΔF refers to a calibrated flow rate (subtracting the $Q_{background}$ flow rate 1.1 SLPM from total flow rate), *K* is the flowrate constant (0.256

SLPM/W in our setup). Thus, a terminal resistance of 0.243 mΩ was calculated and need to be calibrated out by Equation 4.

$$Q_{terminal} = 2.43 \times 10^{-4} I^2 t \tag{4}$$

Where $Q_{terminal}$ is Joule heat by terminal resistance in duration of *t*, *I* is transport current.

*3.3 $Q_{rotation}$ calibration*

When HTS machine was fully operated, the rotor speed is 300 RPM as well as there is also 40 A peak current in stator. The rotation of the rotors will stir liquid nitrogen in the outer cryostat and change the heat transfer balance, this may cause flow rate to change in measurement chamber. As for stator, currently phase B and phase C is copper coils and current will cause Joule heat, although measurement chamber is made of Tufnol which is good thermal isolation material, but there still some heat may transfer to the inner cryostat and cause flow rate changed.

The rotor speed calibration was operated with open-circuit stator windings, thus there is no current in the stator, the rotor speed was measured between 0 to 300 RPM, we record 9 data as Figure 8 shown, when speed increased from 0 to 300 RPM, flow rate decreased from 1.07 SLPM to 0.83 SLPM. Heat transfer is balanced when rotor speed is steady, when liquid nitrogen was stirred by the rotor, some liquid nitrogen cooling the top plate and the foams, causing system proceed to a new thermal balance condition, with the rotor speed increase, temperature difference between measuring chamber and top plate is decreased and cause a lower background flow rate. Thus, the error cause by rotor speed can be calibrated out by Figure 7.

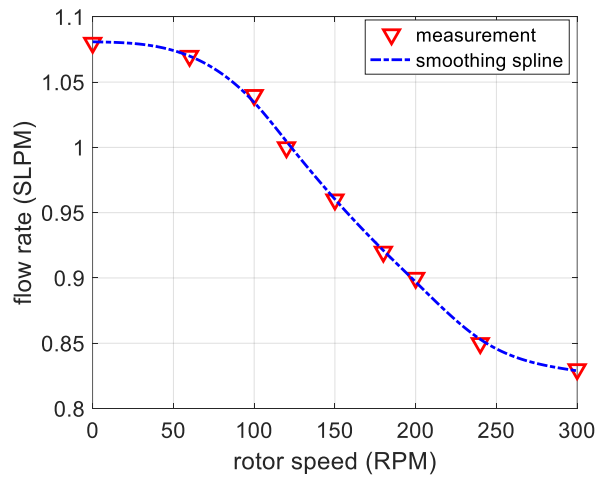

Figure 7. Rotor speed calibration

The stator is designed to operate at a peak current of 40 amps, thus stator calibration is simply kept rotor static, and no input power to inner cryostat, applied currents in all 5 stator coils in outer cryostat, record the flow meter data. The results were shown as Figure 8, the red curve illustrated the DC current applied to the stator, blue line illustrated the flow rate during this time. As we can see from the results that flow rate do not change when different current applied in the stator, Figure 8 proves that good thermal isolation of the Tufnol material and the transfer heat is very small, resulting no influence to the flow rate from stator's Joule heat. Thus, $Q_{rotation}$ is only caused by rotor speed and can be calibrated out by Figure 7.

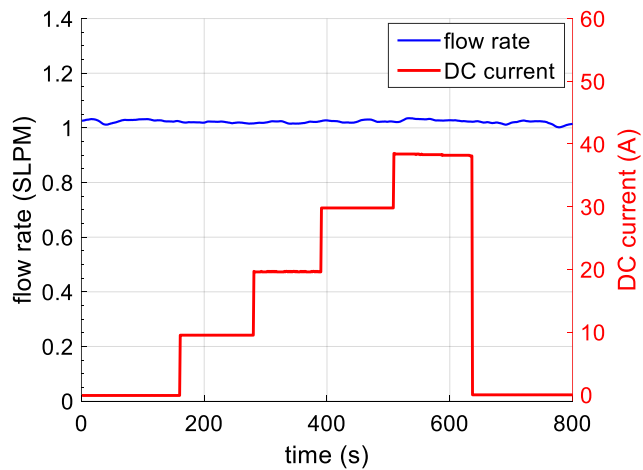

Figure 8. Stator coils current calibration

*3.4 The measurement chamber calibration*

Theoretically, flow rate constant equals to 0.25 standard litre per min per watt (SLPM/W), a resistance wire of 26.4 ohms (in liquid nitrogen) was placed in the measurement chamber and connected to a DC power source. In the heater calibration procedure, various of DC voltages from 0 - 25.2 V was applied in the experiments, 8 points were selected from 3.38 watts to 24.16 watts. Figure 9 (a) demonstrated the results of flow rate (subtracting the background flow) induced by carious heating powers applied to resistance wire. Red curve shows the heater power level and blue curve shows the flow rate after calibrated. Plots these data in one figure as shown in Figure 9 (b), the results show the flow rate proportional to heating power in the measurement chamber, the curve give a ratio of 0.256 SLPM/W refers to flow rate constant, these experiment results show that the actual performance is accordant with the theoretical value with only 2.4% error. On the other hand, Figure 9 validates the system that there was no gas leakage in the measurement chamber. As the fluctuate range of the flow meter is less than 0.08 SLPM which gave a measuring error up to 0.3 W.

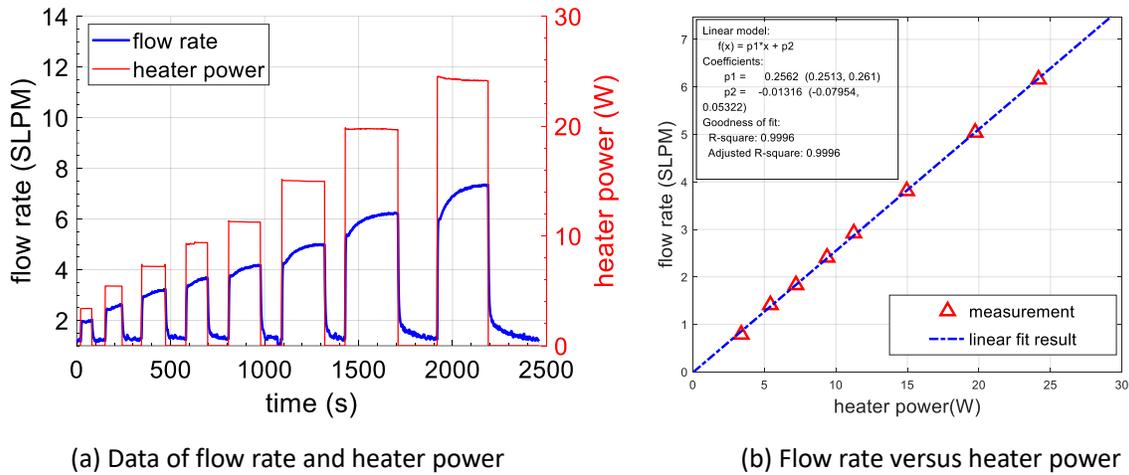

(a) Data of flow rate and heater power  (b) Flow rate versus heater power

Figure 9. Background flow rate data

To sum up, all the errors been estimated and well calibrated, the sensitivity of this testing rig is near 0.3 W, this value is making sense in platform based on calorimetric method.

## 4. AC loss of a HTS stator

The HTS coil is prepared using 25 meters 4mm superpower SCS 4050-AP tape, the single tape critical current is 140 A. The specification of HTS double pancake coil is shown in Table 1, the picture of coil and setup in illustrated in Figure 5 (b).

Table 1. Specification of HTS stator winding coil

| Parameters | value |
| --- | --- |
| Tape type | Superpower SCS4050-AP |
| Tape Ic | 140A |
| Coil Ic (self-field) | 72A |
| Coil inner diameter | 95mm |
| Coil outer diameter | 99.8mm |
| Turns per layer | 38 |
| Total coil turns | 76 |
| inductance | 937.4µH |

Both self-field coil critical current and in field critical current were measured and the results were shown as Figure 10, by using 0.1 µV/cm criteria, the self-filed critical current is 72 A and in field critical current is 53 A (at peak 0.45 T). Thus, this coil is satisfied to operate at peak current of 40 A.

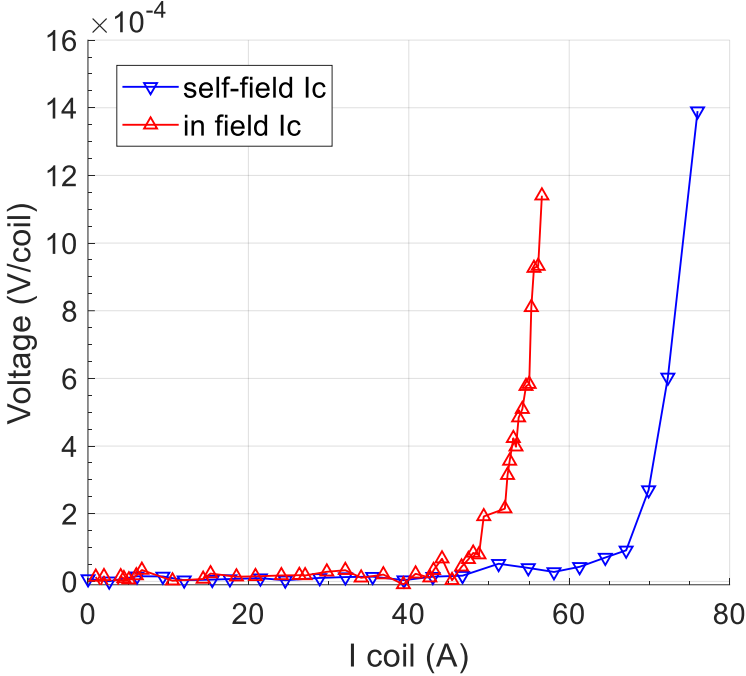

Figure 10. Self-field and in-field critical current

*4.1 magnetisation loss*

The magnetisation loss of this coil can be measured by keeping the HTS coil open circuited, so no transport current in the HTS coil and only the magnetisation loss is measured. The rotor speed was 100, 150,200,240,300 RPM respectively, as this is a two-pole-pair rotor, these speed means the operated frequency with 3.33 Hz, 5 Hz, 6.67 Hz, 8 Hz, 10 Hz respectively. Recording the flow rate of the flow meter during these speeds, the results were shown in Figure 11 (a): blue line illustrated the rotor speed and the red line showed the flow rate. After flow rate calibrated by Equation 3, the magnetisation loss of this coil versus frequency can be calculated by Equation 2 as Figure 12 (b) shown.

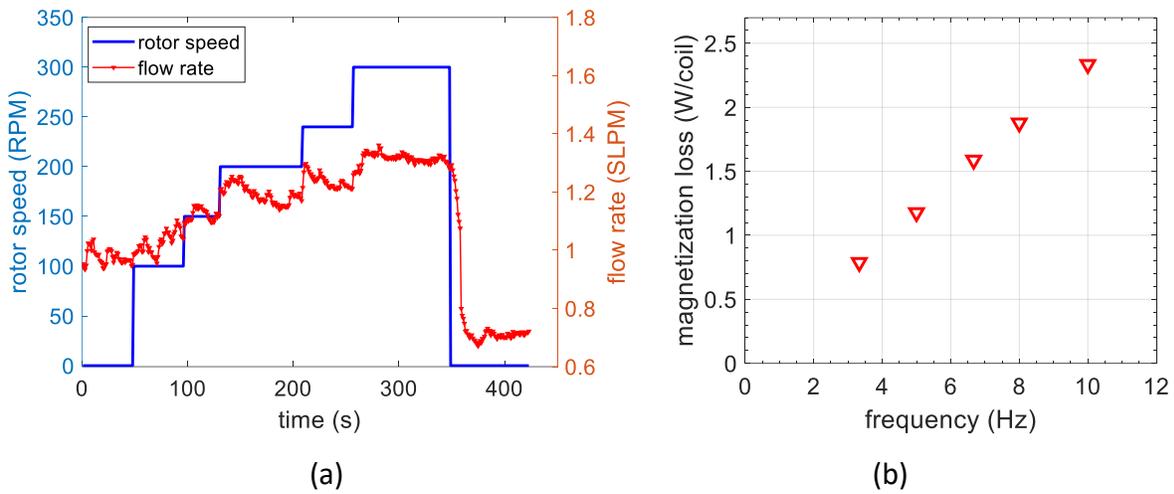

(a)                                    (b)

Figure 11. Magnetisation loss versus frequency

*4.2. total AC loss*

The total AC loss of the single HTS coil is measured by connecting the three-phase stator windings to adjustable power resistors. To measure the total AC loss with various currents, the resistors need to be changed. We measured six different peak transport currents between 10 A and 40 A for four frequencies (3.33 Hz, 5 Hz, 6.67 Hz and 10 Hz), corresponding to rotor speeds at 100, 150, 200 and 300 RPM respectively. Flow rate data from experiments were illustrated in Figure 12, where the blue curves refers to the rotor speed and the red curves refers to instantaneous flow rate value. The transport current values for different frequency are shown in Table 2. The total AC loss versus transport current are shown in Figure 13. The dominated loss in the HTS coil is hysteresis losses at low frequencies, as the losses are largely frequency independent.

Our results show that at a peak current of 73% *Ic* and a peak rotational field of 0.45 T, the HTS coil generates in total 4.56 W of heat at 10 Hz and 48 A. To further understand the impact of a rotational magnetic field, we compared the transport loss of the HTS coil with the total AC loss, as shown in Figure 14. The transport loss was measured in 77 K using the four-point electrical method [19]. The total AC loss is very high for standard 4 mm HTS conductions used as machine windings. To minimize the total AC losses from HTS stator windings, new HTS winding solutions to reduce AC loss are essential. One advantage of this platform is that it enables comparison of the AC loss of different HTS windings. For example, we plan to measure a 1 mm wide multi-filament HTS cable coil with the same geometry as the HTS coil used in this paper and quantify its loss reduction in a machine environment [27, 28].

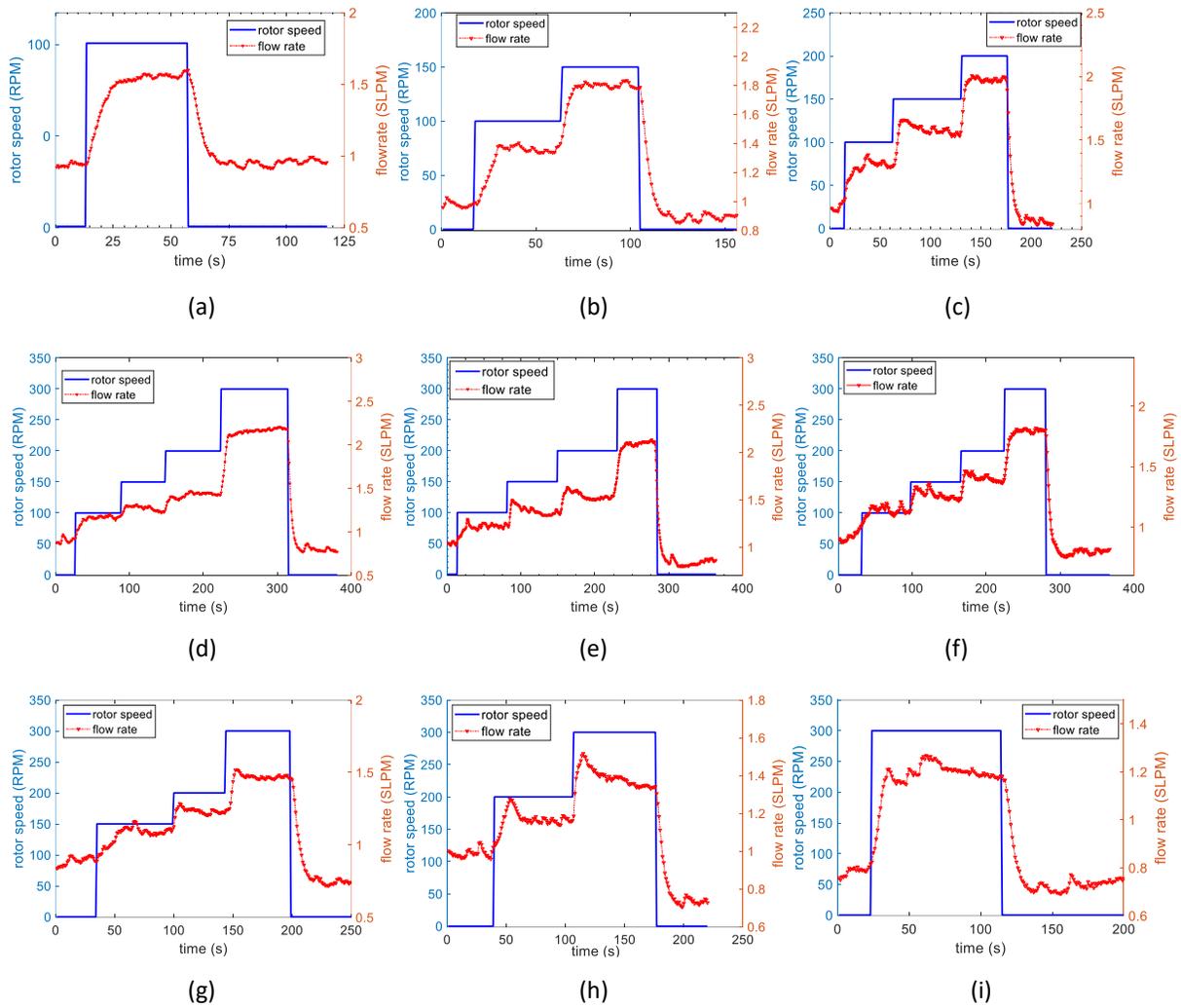

Figure 12. Total AC loss measurements

**Table 2.** Transport current values in Figure 12

|  | 3.33Hz | 5.00Hz | 6.67Hz | 10.00Hz |
|---|---|---|---|---|
| Figure 12 (a) | 40A | | | |
| (b) | 29A | 43.5A | | |
| (c) | 21A | 32A | 41A | |
| (d) | 16A | 24A | 32A | 48A |
| (e) | 14A | 21A | 28A | 42A |
| (f) | 11A | 18A | 24A | 36A |
| (g) | | 11A | 15A | 22A |
| (h) | | | 12A | 18A |
| (i) | | | | 10A |

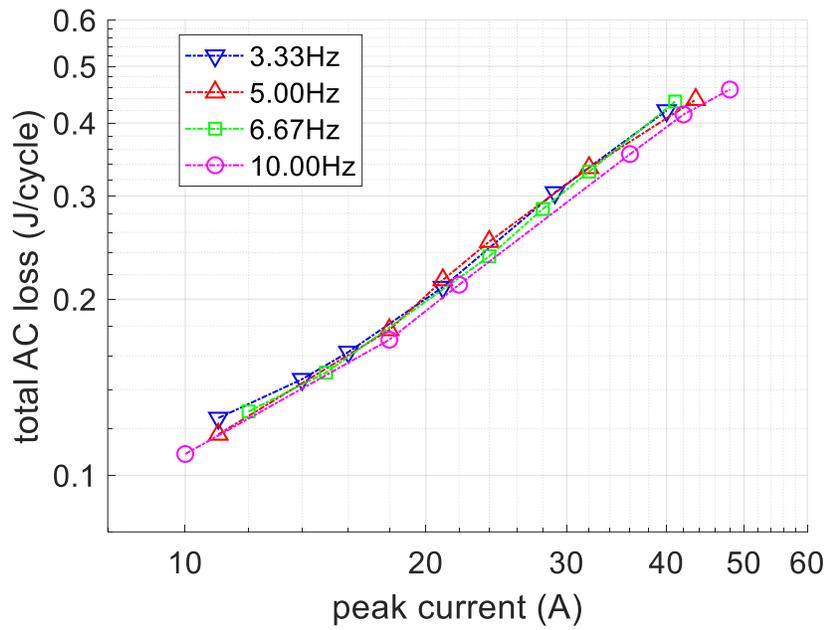

Figure 13. Measured total AC loss

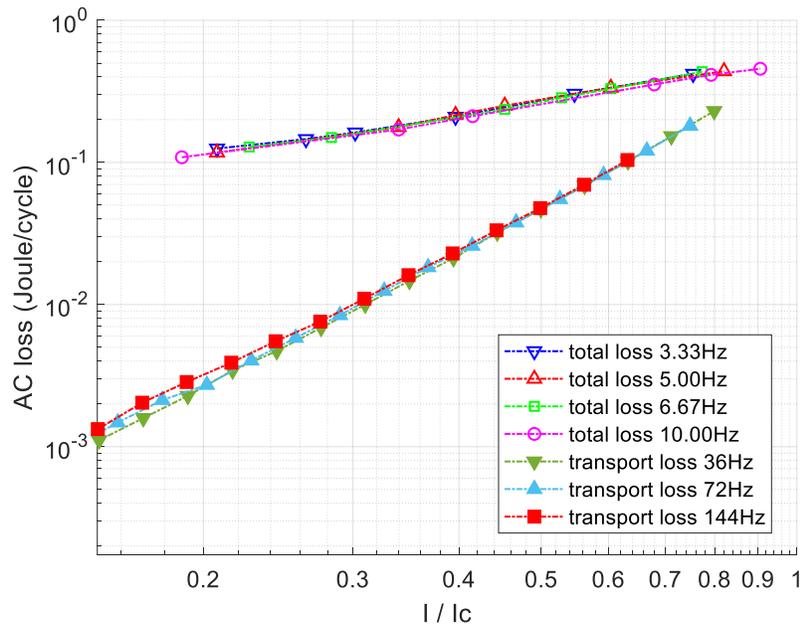

Figure 14. Transport loss and total loss

## 5. Conclusion and future work

The measurement for total AC losses in Figure 13 was performed in a generator mode, so the transport current and the rotational magnetic field are in phase. According to our previous study, the phase angle between the transport current and the external magnetic field affects the total AC losses [6]. When the current and magnetic field are in phase, the total AC loss is the highest. Our next step will be using a machine drive to operate the machine in a motor operational mode so we can control the phase angle and measure its impact to the total AC losses. There is also a possibility to test the impact of power electronic devices to the AC losses of HTS stator by connecting them together and measuring AC losses.

Currently the system works at 77 K by LN2 cooling, but we are developing a helium gas circulation system and a new measurement chamber to enable the total AC losses measurement between 25 K and 77 K, as illustrated in Figure 15. The measurement of AC losses is based on the calorimetric method proposed in [29, 30] by measuring the temperature difference of inlet and outlet helium gas. With our plan to connect the system to a helium gas circulation cooling system, the platform can further be used to characterising different types of superconducting winding, e.g. B2212 and MgB2.

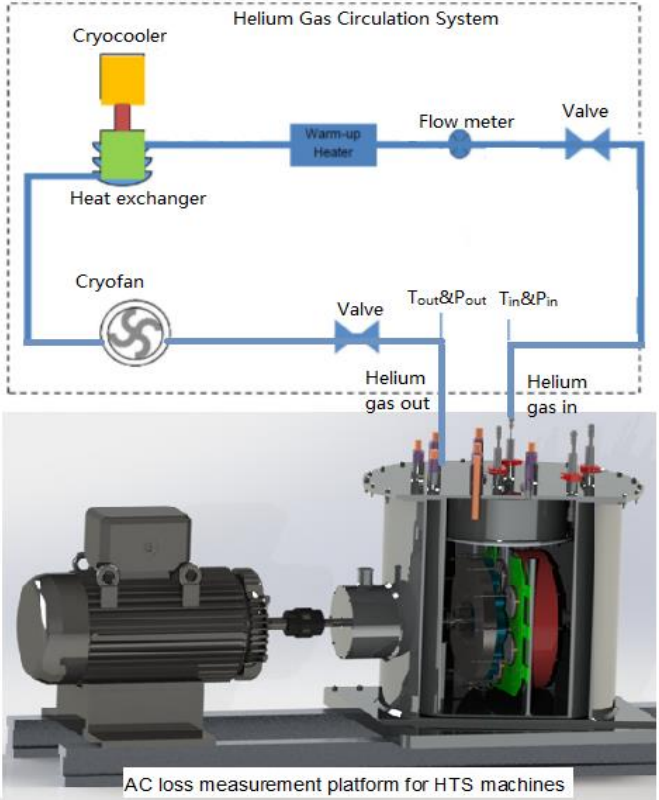

Figure 15. Machine with helium gas circulation system

This paper reports a pioneering testing platform for fully HTS machines used for future electric aircraft machines. The system provides a machine environment to measure AC loss of HTS windings, this systems was carefully calibrated and validated. Focusing on calorimetrically quantify the electrical HTS stator, the platform will provide valuable insights to the AC losses of HTS stator in a rotational machine environment. The platform can be used to the identification of AC loss reduction technologies, contributing to the development of highly efficient fully HTS propulsion machine.

**Acknowledgments**

The project is funded by Royal Academy Research Fellowship under the project "Fully superconducting machines for next generation electric aircraft propulsion" and EPSRC Grant EP/P002277/1 "Developing Highly efficient HTS AC windings for fully superconducting machines".  Fangjing Weng would like to thank the support of Chinese Scholar Council Grant #201908060381. The authors would like to thank Dr Jay Patel in preparing Figure 1 and 3.

# References

[1] Luongo C A, Masson P J, Nam T, Mavris D, Kim H D, Brown G V, Waters M and Hall D 2009 Next generation more-electric aircraft: a potential application for HTS superconductors *IEEE Transactions on Applied Superconductivity* **19** 1055-68
[2] Armstrong M J, Ross C A, Blackwelder M J and Rajashekara K 2012 Propulsion system component considerations for NASA N3-X turboelectric distributed propulsion system *SAE International Journal of Aerospace* **5** 344-53
[3] Sarlioglu B and Morris C T 2015 More electric aircraft: Review, challenges, and opportunities for commercial transport aircraft *IEEE transactions on Transportation Electrification* **1** 54-64
[4] Bertola L, Cox T, Wheeler P, Garvey S and Morvan H 2016 Superconducting Electromagnetic Launch System for Civil Aircraft *IEEE Transactions on Applied Superconductivity* **26**
[5] Berg F, Palmer J, Miller P, Husband M and Dodds G 2015 HTS electrical system for a distributed propulsion aircraft *IEEE Transactions on Applied Superconductivity* **25** 1-5
[6] Brown G 2011 Weights and efficiencies of electric components of a turboelectric aircraft propulsion system. In: *49th AIAA aerospace sciences meeting including the new horizons forum and aerospace exposition,* p 225
[7] Gohardani A S, Doulgeris G and Singh R 2011 Challenges of future aircraft propulsion: A review of distributed propulsion technology and its potential application for the all electric commercial aircraft *Progress in Aerospace Sciences* **47** 369-91
[8] Henke M, Narjes G, Hoffmann J, Wohlers C, Urbanek S, Heister C, Steinbrink J, Canders W R and Ponick B 2018 Challenges and Opportunities of Very Light High-Performance Electric Drives for Aviation *Energies* **11**
[9] Vepa R 2018 Modeling and Dynamics of HTS Motors for Aircraft Electric Propulsion *Aerospace* **5**
[10] Zhang M, Yuan W, Kvitkovic J and Pamidi S 2015 Total AC loss study of 2G HTS coils for fully HTS machine applications *Superconductor Science & Technology* **28** 115011
[11] Ainslie M, Izumi M and Miki M 2016 Recent advances in superconducting rotating machines: an introduction to the 'Focus on Superconducting Rotating Machines' *Superconductor Science & Technology* **29** 060303
[12] Song P, Qu T, Lai L, Wu M, Yu X and Han Z 2016 Thermal analysis for the HTS stator consisting of HTS armature windings and an iron core for a 2.5 kW HTS generator *Superconductor Science & Technology* **29** 054007
[13] Eckels P and Snitchler G 2005 5 MW high temperature superconductor ship propulsion motor design and test results *Naval Engineers Journal* **117** 31-6
[14] Nick W, Grundmann J and Frauenhofer J 2012 Test results from Siemens low-speed, high-torque HTS machine and description of further steps towards commercialisation of HTS machines *Physica C: Superconductivity its applications* **482** 105-10
[15] Li S, Fan Y, Fang J, Qin W, Lv G and Li J 2013 HTS axial flux induction motor with analytic and FEA modeling *Physica C: Superconductivity* **494** 230-4
[16] Wang Y, Song H, Yuan W, Jin Z and Hong Z 2017 Ramping turn-to-turn loss and magnetization loss of a No-Insulation (RE) Ba2Cu3Ox high temperature superconductor pancake coil *Journal of applied physics* **121** 113903
[17] Wang Y, Zhang M, Yuan W, Hong Z, Jin Z and Song H 2017 Non-uniform ramping losses and thermal optimization with turn-to-turn resistivity grading in a (RE) Ba2Cu3Ox magnet consisting of multiple no-insulation pancake coils *Journal of applied physics* **122** 053902
[18] Wang Y, Guan X and Dai J 2014 Review of AC loss measuring methods for HTS tape and unit *IEEE Transactions on Applied Superconductivity* **24** 1-6
[19] Kim J-H, Kim C H, Iyyani G, Kvitkovic J and Pamidi S 2011 Transport AC loss measurements in superconducting coils *IEEE Transactions on Applied Superconductivity* **21** 3269-72
[20] Gömöry F, Vojenčiak M, Pardo E, Solovyov M and Šouc J 2010 AC losses in coated conductors *Superconductor Science & Technology* **23** 034012
[21] Šouc J, Pardo E, Vojenčiak M and Gömöry F 2008 Theoretical and experimental study of AC loss in high temperature superconductor single pancake coils *Superconductor Science & Technology* **22** 015006
[22] Grilli F and Ashworth S P 2007 Measuring transport AC losses in YBCO-coated conductor coils *Superconductor Science & Technology* **20** 794
[23] Zhang M, Wang W, Huang Z, Baghdadi M, Yuan W, Kvitkovic J, Pamidi S and Coombs T 2014 AC loss measurements for 2G HTS racetrack coils with heat-shrink tube insulation *IEEE Transactions on Applied Superconductivity* **24** 1-4
[24] Pardo E 2013 Calculation of AC loss in coated conductor coils with a large number of turns *Superconductor Science & Technology* **26** 105017
[25] Šouc J, Gömöry F and Vojenčiak M 2005 Calibration free method for measurement of the AC magnetization loss *Superconductor Science & Technology* **18** 592
[26] Ekin J 2006 *Experimental techniques for low-temperature measurements: cryostat design, material properties and superconductor critical-current testing*: Oxford university press)
[27] Wang M, Zhang M, Song M, Li Z, Dong F, Hong Z and Jin Z 2018 An effective way to reduce AC loss of second-generation high temperature superconductors *Superconductor Science & Technology* **32** 01LT
[28] Grilli F and Kario A 2016 How filaments can reduce AC losses in HTS coated conductors: a review *Superconductor Science & Technology* **29** 083002
[29] Kvitkovic J, Hatwar R and Pamidi S 2016 Simultaneous magnetic shielding and magnetization loss measurements of YBCO cylinders at variable temperatures under cryogenic helium gas circulation *IEEE Transactions on Applied Superconductivity* **26** 1-5
[30] Van der Laan D, Weiss J, Kim C, Graber L and Pamidi S 2018 Development of CORC® cables for helium gas cooled power transmission and fault current limiting applications *Superconductor Science & Technology* **31** 085011